\documentclass[pra,twocolumn,showpacs,amsmath,amssymb]{revtex4}
\usepackage{graphicx}
\usepackage{bm}
\usepackage{amsmath,amssymb}

\begin{document}

\title{Vacuum entanglement governs the bosonic character of magnons}  

\author{Tomoyuki Morimae}
\email{morimae@gmail.com}
           \affiliation{
Laboratoire Paul Painlev\'e, Universit\'e Lille 1,
F-59655 Villeneuve d'Ascq Cedex, France
}
\date{\today}
            
\begin{abstract}
It is well known that magnons, elementary excitations in a magnetic material,
behave as bosons when their density is low.
We study how the bosonic character of magnons
is governed by the amount of a multipartite entanglement in the vacuum
state on which magnons are excited.
We show that if the multipartite entanglement is strong, 
magnons cease to be bosons.
We also consider some examples,
such as ground states of the Heisenberg ferromagnet and the transverse Ising model, the condensation of magnons,
the one-way quantum computer, and Kitaev's toric code.
Our result provides insights into the quantum statistics of elementary
excitations in these models, and into the reason why a non-local transformation,
such as the Jordan-Wigner transformation, is necessary for some many-body systems. 
\end{abstract}
\pacs{03.65.Ud, 03.67.Bg, 03.67.Mn}
\maketitle  
\section{Introduction}
The study of condensed matter physics
from the view point of quantum information science~\cite{Nielsen}
is one of the most active research topics in today's quantum many-body 
physics.
Plenty of important results have been obtained
and they have contributed to the recent mutual fertilization
between those two fields~\cite{Amico,Guhne,Horodecki}.
Among many characteristics of quantum many-body systems,
ground states and low-energy excited states are 
fundamental targets in condensed matter physics~\cite{Mahan}.
Indeed,
a great deal of research has been performed
on the behavior of an entanglement in ground states~\cite{Latorre}, 
and these
researches have revealed several novel aspects of 
quantum ground states including
the entanglement divergence at a quantum critical point~\cite{ent_div} and
the role of an entanglement in 
efficient approximations of many-body ground states~\cite{MPS}.

Although ground states of many-body Hamiltonians 
are often very complicated,  
it is notorious that excited states are more complicated due to 
their high degeneracy.
Because of this complexity of excited states, the
study of excited states from the view point of quantum information is 
very slow and not yet fully developed.
This seems to cause the severe lack of knowledge about the implementation
of quantum computation
on low-temperature solids.
In the long history of condensed matter physics,
one solution to cope with such complicated
excited states
has been the idea of
``elementary excitations"~\cite{Nakajima}.
A low-energy excited state of a many-body Hamiltonian
can be interpreted as a ripple of the degrees of freedom
under consideration
(such as the electromagnetic field
or the density of atoms),
and this ripple is second quantized
to have (quasi-)particles (such as photons or phonons) 
whose lifetimes are not necessarily infinite.   
Such particles are called 
``elementary excitations"~\cite{Nakajima}.
For a magnetic material, such as a ferromagnet or an antiferromagnet,
such a quantum of a spin wave
is called ``magnon". 
Low-energy excited states of a magnetic Hamiltonian are
labelled with the density and wavenumbers of magnons excited on the vacuum,
and this labelling enables a systematic study of otherwise 
highly complicated low-energy excited states.
Indeed,
this ``magnon picture" 
has been applied to several researches in quantum information, such as 
entanglement properties of spin systems~\cite{M,Illuminati}
and the transfer of quantum information through a spin chain~\cite{transfer}.
Considering facts that elementary excitations are at the
heart of quantum many-body physics, and that a quantum information
processing will be ultimately implemented on a quantum many-body system,
we cannot overestimate
the role of elementary excitations played in
quantum information science. 

Recently, Law~\cite{Law}
and Chudzicki, Oke, and Wootters~\cite{Wootters} have 
shown
an interesting relation between
the bosonic character of composite fermions 
and
the entanglement between these fermions.
Their main result is
that if the entanglement is strong, an
indicator of the bosonic character
of the composite fermions
approaches its ideal bosonic value.
In other words, the composite fermions behave as bosons,
when the entanglement is strong.
Their researches have opened the door to the
study of the quantum statistics from the view point of
entanglement.

In this Rapid Communication,
we study how the bosonic character of magnons
is governed by the amount of a multipartite entanglement 
in the vacuum state
on which magnons are excited.
Our main result is Eq.~(\ref{result}), which shows that 
if the multipartite entanglement is strong, 
magnons cease to be bosons.

One might claim that our result ``strong entanglement means weak bosonic character"
is contradictory to the result ``strong entanglement means strong bosonic character" by Law~\cite{Law}, Chudzicki, Oke, and Wootters~\cite{Wootters}.
However, note that the entanglement we consider here is different from that considered by them: 
we see a multipartite entanglement in the vacuum on which particles are
excited, whereas they see the bipartite 
entanglement between two excited particles.
Of course, it would be interesting to explore a relation between our result
and theirs.
It is, however, beyond the scope of the present paper, and therefore left for a future study.

As examples, we consider several typical
quantum many-body systems, such as
the Heisenberg ferromagnet, the transverse Ising model,
the one-way quantum computer~\cite{one-way}, 
and Kitaev's toric code~\cite{Kitaev}.
We will see that our result Eq.~(\ref{result})
provides an insight into the quantum statistics of elementary
excitations in these models.
We will also see that
our result explains why a {\it non-local} transformation, such
as the Jordan-Wigner transformation~\cite{Sachdev}, 
is necessary for some many-body systems, such as the transverse Ising
model.
Furthermore, we also consider the condensation of magnons. Since
a state where a macroscopic number of 
magnons are excited is highly entangled~\cite{M},
Eq.~(\ref{result}) also explains the 
often-pointed-out fact that 
magnons cease to be bosons
when their density is high.

We would like to emphasize that
the result of this Rapid Communication sublimates plenty of recent calculations 
of the multipartite entanglement
in ground states from mere academic studies into
meaningful applications in condensed matter physics:
calculations of the multipartite entanglement in ground states
are useful to obtain an insight into the quantum statistics of elementary excitations
excited there.

\section{Magnon}
For simplicity, let us consider the one-dimensional lattice
of $N$ spin-1/2 particles.
The creation operator $\hat{M}_k^\dagger$ of the magnon with wavenumber $k$
is defined by 
\begin{eqnarray*}
\hat{M}_k^\dagger\equiv\frac{1}{\sqrt{N}}\sum_{l=1}^N e^{ikl}\hat{\sigma}_l^+,
\end{eqnarray*}
where 
$\hat{\sigma}_l^+$ is the ``spin-ladder" operator on site $l$ defined by
$\hat{\sigma}_l^+|0\rangle_l=|1\rangle_l$ and $\hat{\sigma}_l^+|1\rangle_l=0$.
Here, $|1\rangle_l$ and $|0\rangle_l$ are eigenvectors of 
Pauli's $z$ operator $\hat{\sigma}_l^z$ on site $l$
corresponding to eigenvalues
$\pm1$, respectively. 
The annihilation operator of the magnon with wavenumber $k$
is the Hermitian conjugate $\hat{M}_k$
of the corresponding creation operator $\hat{M}_k^\dagger$.
It is easy to check that they satisfy
\begin{eqnarray*}
[\hat{M}_k,\hat{M}_{k'}]=
[\hat{M}_k^\dagger,\hat{M}_{k'}^\dagger]=0.
\end{eqnarray*}
Furthermore,
in condensed matter physics, 
it is often said that~\cite{Nakajima}
magnons are approximately bosons
when their density is low.
This is ``intuitively" denoted by
\begin{eqnarray*}
[\hat{M}_k,\hat{M}_{k'}^\dagger]
\simeq \delta_{k,k'},
\end{eqnarray*}
where the mathematically exact meaning of ``$\simeq$" depends on the context.
Low-energy excited states of magnetic Hamiltonians
are often approximated very well by few-magnon states, such as
$\hat{M}_{k_1}^\dagger|\psi\rangle$
and
$\hat{M}_{k_1}^\dagger\hat{M}_{k_2}^\dagger|\psi\rangle$
with an appropriate vacuum state $|\psi\rangle$~\cite{Mahan}.
Furthermore, a state where  
a macroscopic number of magnons are condensed is 
often used to describe the behavior of a macroscopic order
when the many-body system exhibits a phase transition~\cite{Nakajima}.
The condensation of magnons has been
experimentally observed~\cite{BECmagnon1}.

\section{Bosonic character and entanglement}
In Refs.~\cite{Law,Wootters}, 
the quantity
$\langle n|[\hat{c},\hat{c}^\dagger]|n\rangle$
was used as an indicator of the bosonic character of composite fermions,
where $\hat{c}^\dagger$ and $\hat{c}$ are creation and annihilation
operators of the composite fermions,
respectively,
and $|n\rangle$ is the state on which $n$ composite fermions are excited.
Here, we consider the similar quantity
$|\langle\psi|[\hat{M}_k,\hat{M}_k^\dagger]|\psi\rangle|$
as the indicator of the bosonic character of magnons.
This quantity should be 1 if magnons are exactly bosons.
However, this quantity is upper bounded as~\cite{kk}
\begin{eqnarray}
|\langle\psi|[\hat{M}_k,\hat{M}_k^\dagger]|\psi\rangle|
&=&
\frac{1}{N}
\Big|
\sum_{l=1}^N
\langle\psi|[\hat{\sigma}_l^-,\hat{\sigma}_l^+]|\psi\rangle
\Big|\nonumber\\
&\le&
\frac{1}{N}
\sum_{l=1}^N
|\langle\psi|\hat{\sigma}_l^z|\psi\rangle|\nonumber\\
&\le&
\frac{1}{N}
\sum_{l=1}^N
\sqrt{1-S_l}\equiv\Lambda,
\label{result}
\end{eqnarray}
where 
$S_l\equiv2[1-\mbox{Tr}(\hat{\rho}_l^2)]$
is the entanglement between $l$th spin and other spins~\cite{Law,Wootters},
$\hat{\rho}_l\equiv\mbox{Tr}_l(|\psi\rangle\langle\psi|)$
is the reduced density operator for $l$th spin,
$\mbox{Tr}_l$ means the trace over all spins
except for $l$th spin,
and we have used
the relation~\cite{relation}
\begin{eqnarray*}
1-S_l&=&
\langle\psi|\hat{\sigma}_l^x|\psi\rangle^2
+\langle\psi|\hat{\sigma}_l^y|\psi\rangle^2
+\langle\psi|\hat{\sigma}_l^z|\psi\rangle^2\\
&\ge&
\langle\psi|\hat{\sigma}_l^z|\psi\rangle^2.
\end{eqnarray*}
Here $\hat{\sigma}_l^x$
and $\hat{\sigma}_l^y$
are Pauli's $x$ and $y$ operators on site $l$, respectively.
$S_l$ satisfies $0\le S_l \le 1$.
If $l$th spin is maximally entangled with other spins, $S_l=1$.
On the other hand,
if $l$th spin is separable from other spins, $S_l=0$. 
The physical meaning of Eq.~(\ref{result}) is that
if the multipartite entanglement in $|\psi\rangle$ is strong
in the sense that $S_l$ is close to 1 for many $l$'s,
then 
$|\langle\psi|[\hat{M}_k,\hat{M}_k^\dagger]|\psi\rangle|
\ll 1$,
which means magnons are not bosons.

\section{Examples}
Let us investigate consequences of Eq.~(\ref{result}) by studying some examples. First,
we consider
the Heisenberg ferromagnet: 
\begin{eqnarray*}
-\sum_{l=1}^N [
\hat{\sigma}_l^x\hat{\sigma}_{l+1}^x
+\hat{\sigma}_l^y\hat{\sigma}_{l+1}^y
+\hat{\sigma}_l^z\hat{\sigma}_{l+1}^z
].
\end{eqnarray*}
A ground state
of this Hamiltonian is $\bigotimes_{l=1}^N|0\rangle_l$~\cite{rotation},
and therefore
we obtain $S_l=0$ for all $l$. This means
$\Lambda=1$ and Eq.~(\ref{result}) does not
prohibit magnons from being bosons.
Indeed, 
the Heisenberg ferromagnet is a canonical example in the statistical physics
where few magnons on the ground state 
behave as bosons~\cite{Nakajima,HF}.

Second, let us consider the condensation of magnons.
The state where $m$ magnons are condensed in wavenumber 0
is the $m$-Dicke state:
\begin{eqnarray*}
\Big(\sum_{l=1}^N\hat{\sigma}_l^+\Big)^m
\bigotimes_{l=1}^N|0\rangle_l.
\end{eqnarray*}
Let $m=\alpha N$, where $0< \alpha \le1/2$.
It is easy to show that $S_l=4\alpha(1-\alpha)$ for all $l$.
Therefore, $\Lambda=1-2\alpha<1$. 
(In particular, if $\alpha=1/2$, $\Lambda=0$.)
This means that if a macroscopic number of magnons are condensed, 
they do not behave as bosons.
Indeed, it is often pointed out that magnons cease to be bosons
if their density is high~\cite{Nakajima}.

Third, let us consider 
the one-way quantum computer~\cite{one-way}.
The one-way quantum computation is performed by adaptively measuring
each qubit in the cluster state, which is defined 
as the simultaneous
eigenvector of operators $\hat{K}_l$ $(l=1,2,...,N)$:
\begin{eqnarray*}
\hat{K}_l\equiv\hat{\sigma}_l^x\prod_{m\in C(l)}\hat{\sigma}_m^z
\end{eqnarray*}
corresponding to the simultaneous eigenvalue 1,
where $C(l)$ is the set of nearest-neighbor sites
of $l$.
To analyze the behavior of elementary excitations on the cluster state
is important from the view point of the fault-tolerant one-way quantum computation,
since the propagation of an error is in the shape of elementary excitations.
It is easy to understand that $S_l>0$ for all $l$ in the cluster state,
since otherwise a single site is separable from other sites
and therefore a measurement on this site
does not contribute to the quantum computation.
In fact, it can be shown by a direct calculation that $S_l=1$ for 
all $l$~\cite{one-way}. 
Therefore, we conclude that magnons on the cluster state are not bosons.

Fourth, let us consider
the transverse Ising model:
\begin{eqnarray*}
-\sum_{l=1}^N [
\hat{\sigma}_l^z\hat{\sigma}_{l+1}^z
+B\hat{\sigma}_l^x
],
\end{eqnarray*}
where $B$ is the transverse magnetic field.
This model exhibits a quantum phase transition at $B=1$~\cite{Sachdev}.
When $0<B\ll1$, the ground state is approximately~\cite{infact} the $N$-qubit 
Greenberger-Horne-Zeilinger (GHZ) state
$\frac{1}{\sqrt{2}}(
|0^{\otimes N}\rangle+|1^{\otimes N}\rangle
)$~\cite{symmetrybreaking}.
Then, it is obvious that
$S_l=1$ for all $l$'s.
Therefore, $\Lambda=0$ and this means magnons are not bosons.
(Indeed, it is well-known that the transverse Ising model is 
well described by a set of fermions through the Jordan-Wigner 
transformation~\cite{Sachdev}.)
Note that even if we redefine the creation operator of a magnon by
using another local operator $\hat{a}_l^\dagger$ as
\begin{eqnarray*}
\frac{1}{\sqrt{N}}\sum_{l=1}^N e^{ikl}\hat{a}_l^\dagger,
\end{eqnarray*}
``magnons" defined in this way are still not bosons, since
the quantity
$|\langle\psi|[\hat{a}_l,\hat{a}_l^\dagger]|\psi\rangle|$,
which is the expectation value of a local operator on site $l$,
is small if 
the mixedness of the reduced density operator $\hat{\rho}_l$ of site $l$ is strong.
This is one explanation why a {\it non-local} transformation, such as
the Jordan-Wigner transformation, is necessary for the
transverse Ising model.

Fifth, we can also consider more complicated ground states.
For example, in Ref.~\cite{Latorre}, 
the block entanglement in the ground state
of the XY model was calculated.
According to their result, $S_l>0$ 
for various parameters,  
and therefore Eq.~(\ref{result}) suggests
that magnons are not bosons.
Indeed, as is explained there, the XY model is
exactly solved by introducing fermions through 
the Wigner-Jordan transformation.

Finally, the stability of quantum error-correcting codes is also important.
For example, let us consider 
Kitaev's toric code~\cite{Kitaev}.
It is easy to understand $S_l>0$ for all $l$, since
otherwise a single site is separable from other sites and therefore
an error on this site cannot be detected by syndrome measurements.
In fact, it is well
known that elementary excitations in the 
toric code are non-local anyons~\cite{Kitaev}.

\section{Conclusion and Discussion}
In this Rapid Communication, we have studied how the multipartite
entanglement in the vacuum state governs
the bosonic character of magnons excited there.
We have shown that if the multipartite entanglement is strong, magnons cease
to be bosons.

As is well known, there are many different kinds of measures for multipartite
entanglement, and each measure sees different aspects of a quantum many-body
system.
What we have considered here is an average of local entanglement.
Such a way of quantifying multipartite entanglement is often adopted by many
researchers.
Among them, the most famous definition would be the ``global entanglement"
\begin{eqnarray*}
G\equiv2\Big[1-\frac{1}{N}\sum_{l=1}^N\mbox{Tr}(\hat{\rho}_l^2)\Big]
\end{eqnarray*}
given by Meyer and Wallach~\cite{Meyer}. 
Here, let us point out
that the plausible relation 
\begin{eqnarray*}
|\langle\psi|[\hat{M}_k,\hat{M}_k^\dagger]|\psi\rangle|\stackrel{?}{\le} 1-G
\end{eqnarray*}
does not hold because of the counter example: 
$|\psi\rangle=\sqrt{\frac{3}{4}}|0^{\otimes N}\rangle+
\sqrt{\frac{1}{4}}|1^{\otimes N}\rangle$,
which gives
$|\langle\psi|[\hat{M}_k,\hat{M}_k^\dagger]|\psi\rangle|
=\frac{1}{2}
>
\frac{1}{4}
= 1-G$.
It would be an interesting subject of future study to clarify which 
types of multipartite entanglement
measures govern the bosonic character of magnons.

An important physical application of our result 
would be the quantum memory~\cite{memory}. 
Quantum information encoded on   
a photon can be mapped to 
the collective excitation of 
atoms,
and this information is later retrieved on demand.
Since this collective excitation is in a shape of magnons,
our result suggests that the ``blank" state of the memory
should not be highly entangled
if one wants to
preserve the bosonic character of magnons 
which carries the quantum information of incoming photons.

Finally, although we have considered a discrete
field (spin lattice) here, 
it would be interesting to generalize our result to a continuous field,
since there are several works studying
the violation of Bell's inequality 
at the vacuum of a continuous field~\cite{cont}.

\acknowledgements
The author acknowledges the support by the French Agence Nationale de la
Recherche (ANR) under the grant StatQuant (JC07 07205763).


\begin{thebibliography}{00}
\bibitem{kk}
We obtain the same inequality even if we consider
$[\hat{M}_k,\hat{M}_{k'}^\dagger]$
since $|e^{-ikl+ik'l'}|=1$.
\bibitem{relation}
The first equality is shown by seeing that
a state $\hat{\rho}_l$ is generally written as
$\hat{\rho}_l=\frac{1}{2}
(\mbox{Tr}(\hat{\rho}_l\hat{\sigma}_l^x)\hat{\sigma}_l^x
+\mbox{Tr}(\hat{\rho}_l\hat{\sigma}_l^y)\hat{\sigma}_l^y
+\mbox{Tr}(\hat{\rho}_l\hat{\sigma}_l^z)\hat{\sigma}_l^z
+\hat{1})$
and therefore the purity is given by
$\mbox{Tr}(\hat{\rho}_l^2)=\frac{1}{2}
(\mbox{Tr}(\hat{\rho}_l\hat{\sigma}_l^x)^2
+\mbox{Tr}(\hat{\rho}_l\hat{\sigma}_l^y)^2
+\mbox{Tr}(\hat{\rho}_l\hat{\sigma}_l^z)^2
+1)$.
\bibitem{rotation}
Because of the rotational symmetry of the Hamiltonian, other product states
where all spins point the same direction are also ground states.
When another ground state than $|0^{\otimes N}\rangle$ 
is considered, the definition of the creation
operator of a magnon should be appropriately modified.
\bibitem{HF}
Indeed, it is easy to confirm that $\hat{M}_k^\dagger\bigotimes_{l=1}^N|0\rangle_l$
is the exact energy eigenstate of the Heisenberg ferromagnet,
and that
$(\bigotimes_{l=1}^N\langle0|_l)
[\hat{M}_k,\hat{M}_{k'}^\dagger]
(\bigotimes_{l=1}^N|0\rangle_l)=\delta_{k,k'}$.
\bibitem{infact}
In fact, the approximation of the ground state
by the GHZ state is not a good one, since
according to the Peron-Frobenius theorem~\cite{matrix}, the ground state of the 
transverse Ising model has no node, that is,
the ground state is a superposition of all $2^N$ $N$-qubit computational basis.
However, since
the contribution of $\bigotimes_{l=1}^N|0\rangle_l$
and $\bigotimes_{l=1}^N|1\rangle_l$ to the superposition
is large compared with that by other computational basis~\cite{tim_morimae},
such approximation is convenient to capture the essence.
In Ref.~\cite{tim_morimae}, it was shown that the exact ground state
of the transverse Ising model for $0<B<1$
has a superposition of macroscopically distinct states.
According to the result of Ref.~\cite{macro_ent}, such macroscopic superposition
means a strong multipartite entanglement in the sense that
$S_l=O(1)$ for a macroscopic number $[O(N)]$ of $l$'s, where
$f(N)=O(N^r)$ means $\lim_{N\to\infty}f(N)/N^r=$const.$\neq0$.
If we consider the periodic boundary condition,
this means $S_l=O(1)$ for all $l$
because of the translational symmetry.
Then, $\Lambda<1$ and therefore we obtain the same conclusion
that magnons are not bosons.
\bibitem{symmetrybreaking}
If a symmetry breaking field is introduced, the symmetry of the ground
state is broken, and one of two separable states 
$|0^{\otimes N}\rangle$ or $|1^{\otimes N}\rangle$
is obtained. In this case, we obtain the bosonic magnon 
since these symmetry-broken ground states are separable.
\end{thebibliography}
\end{document}